\documentclass[9pt,twocolumn,twoside]{osajnl}

\journal{ol} 

\setboolean{shortarticle}{true}
\usepackage{ulem}

\title{Ultra high-Q WGM microspheres from ZBLAN for the mid-IR band}

\author[1,2,*]{Tatiana S. Tebeneva}
\author[1]{Artem E. Shitikov}
\author[2,3]{Oleg V. Benderov}
\author[1]{Valery E. Lobanov}
\author[1,4]{Igor A. Bilenko}
\author[2]{Alexander V. Rodin}

\affil[1]{Russian Quantum Center, 143026 Skolkovo, Russia}
\affil[2]{Moscow Institute of Physics and Technology, 141701, Dolgoprudny, Russia}
\affil[3]{Institute for Theoretical and Applied Electromagnetics, Russian Academy of Science, 125412, Moscow, Russia}
\affil[4]{Faculty of Physics, Lomonosov Moscow State University, 119991 Moscow, Russia}

\affil[*]{Corresponding author: tetasia19@gmail.com}

\begin{abstract}

The advantages of high-quality-factor whispering gallery mode microresonators can be applied to develop novel photonic devices for the mid-IR range. ZBLAN (glass based on heavy metal fluorides) is one of the most promising materials to be used for this purpose due to low optical losses in the mid-IR. We developed original fabrication method based on melting of commercially available ZBLAN-based optical fiber to produce high-Q ZBLAN microspheres with the diameters of 250 to 350 $\mu$m. We effectively excited whispering gallery modes in these microspheres and demonstrated high quality factor both at 1.55 $\mu$m and 2.64 $\mu$m. Intrinsic quality factor at telecom wavelength was shown $(5.4\pm0.4)\cdot10^8$ which is defined by the material losses in ZBLAN. In the mid-IR at 2.64 $\mu$m we demonstrated record quality factor in ZBLAN exceeding $10^8$ which is comparable to the highest values of the Q-factor among all materials in the mid-IR.
\end{abstract}

\begin{document}

\maketitle

\section{Introduction}
Optical microresonators with whispering gallery modes (WGM) combine a small mode volume, high-Q-factors in a wide spectral range \cite{braginsky_quality-factor_1989, vahala_optical_2003} and can be assembled into a compact device \cite{Ilchenko2006-II}. High-Q-factor WGM microresonators with unique properties are of interest for a variety of applications such as biosensing analysis  \cite{Zhu2010,Ward2011}, for realization of different nonlinear effects \cite{Lin:17,Kippenberg2018,PASQUAZI20181}, for creating novel laser sources \cite{deng_demonstration_2014, behzadi_power_2018, he_whispering_2013}, as well as for the stabilization of different lasers \cite{Liang2015,Galiev:18, shitikov2021self}.To date high-Q WGM microresonators made of various materials used for the near-IR photonics (silicon \cite{yu_mode-locked_2016, griffith_silicon_2015, shitikov2020microresonator}, sapphire \cite{shankar_integrated_2013}, silica \cite{braginsky_quality-factor_1989, klitzing_frequency_2001, armani_ultra-high-q_2003, anashkina_optical_2021, anashkina_kerr_2021, shen_observation_2015}, and other crystalline and amorphous materials were demonstrated. Compared to crystalline materials glass WGM microresonators can be easily produced by melting techniques without any additional mechanical polishing required for crystalline microresonators \cite{ilchenko_nonlinear_2004,Fujii:20}. 
In a pioneering work \cite{braginsky_quality-factor_1989} compact high-quality-factor spherical microresonators made of fused silica with a relatively simple manufacturing process have been demonstrated for the first time. Silica glasses have low optical losses in the near-IR range and to date the maximum obtained Q-factors in WGM microresonators made of fused silica exceed $10^9$ at a wavelength of 1.55 $\mu$m \cite{gorodetsky_ultimate_1996}. However, losses in silica rapidly increase due to the infrared absorption from 1.7 $\mu$m \cite{miya1979ultimate}, so that for longer wavelength chalcogenide and fluorozirconate fibers are used. 
The advantages of WGM microresonators make it possible to develop WGM photonics in the range from 2 $\mu$m, but this, among other issues, requires optical materials that are transparent in this range.
One of the suitable material is glass based on heavy metal fluorides ZBLAN (a group of glasses with the composition ZrF$_4$-BaF$_2$-LaF$_3$-AlF$_3$-NaF, discovered in 1974 at the University of Rennes in France \cite{poulain_verres_1975}). ZBLAN glass is transparent in the range from 250 nm to $\sim$4.5 $\mu$m, with a refractive index about 1.5 at the range of 1.5 to 2.7 $\mu$m \cite{zhang1994evaluation} and has low glass transition temperature ($\sim$265 \textdegree C) \cite{harrington_2004}. Since the 1980s, interest to ZBLAN glasses has increased due to the fact that the calculated theoretical minimum loss for them was 0.01 dB/km at a wavelength range of 2–3 $\mu$m \cite{parker_fluoride_1989}. WGM microresonators made of ZBLAN with $Q = 10^7$ at a wavelength of 1.55 $\mu$m were demonstrated in \cite{way_high-q_2012} in 2012. This result was by an order less than expected from the calculated absorption-limited value at 1.55 $\mu$m apparently caused by residual bulk and surface scattering losses. The commercially available fluoride fiber used in \cite{way_high-q_2012} had 10 times bigger attenuation than currently available fibers (\cite{saad_fluoride_2009}).
Since ZBLAN glass is suitable for doping with various rare earth elements and has a small non-linearity coefficient, it is applicable for the development of high-efficient micro-lasers and amplifiers \cite{wetenkamp_optical_1992}. Lasing at a wavelength of 2.71 $\mu$m, with optical pump at 980 nm, was demonstrated in Er:ZBLAN-doped microsphere \cite{deng_demonstration_2014}. \textcolor{black}{It is worth noting that ZBLAN has anomalous material group velocity dispersion from 1.6 $\mu$m \cite{zhang1994evaluation} that could be used in combination with high achievable Q-factor for generation of soliton microcombs (dissipative Kerr solitons) in the mid-IR.}
In our work, we optimized the fabrication process to produce the high-Q-factor ZBLAN microspheres made of corresponding commercially available optical fiber. The fabricated resonators were characterized in the near- and mid-IR ranges. WGM were excited at the at wavelengths of 1.55 and 2.64 $\mu$m with a prism as a coupling element. The intrinsic Q-factor of the excited modes at 1.5 $\mu$m was $(5.4\pm0.4)\cdot10^8$ which is close to the absorption limit of the commercial ZBLAN fibers. For the 2.64 $\mu$m Q-factor was measured as $(1.13\pm0.22)\cdot10^8$. 

\section{Fabrication of microspherical WGM ZBLAN resonators}
\begin{figure}[htbp!]
\centering
\includegraphics[width=0.55\linewidth]{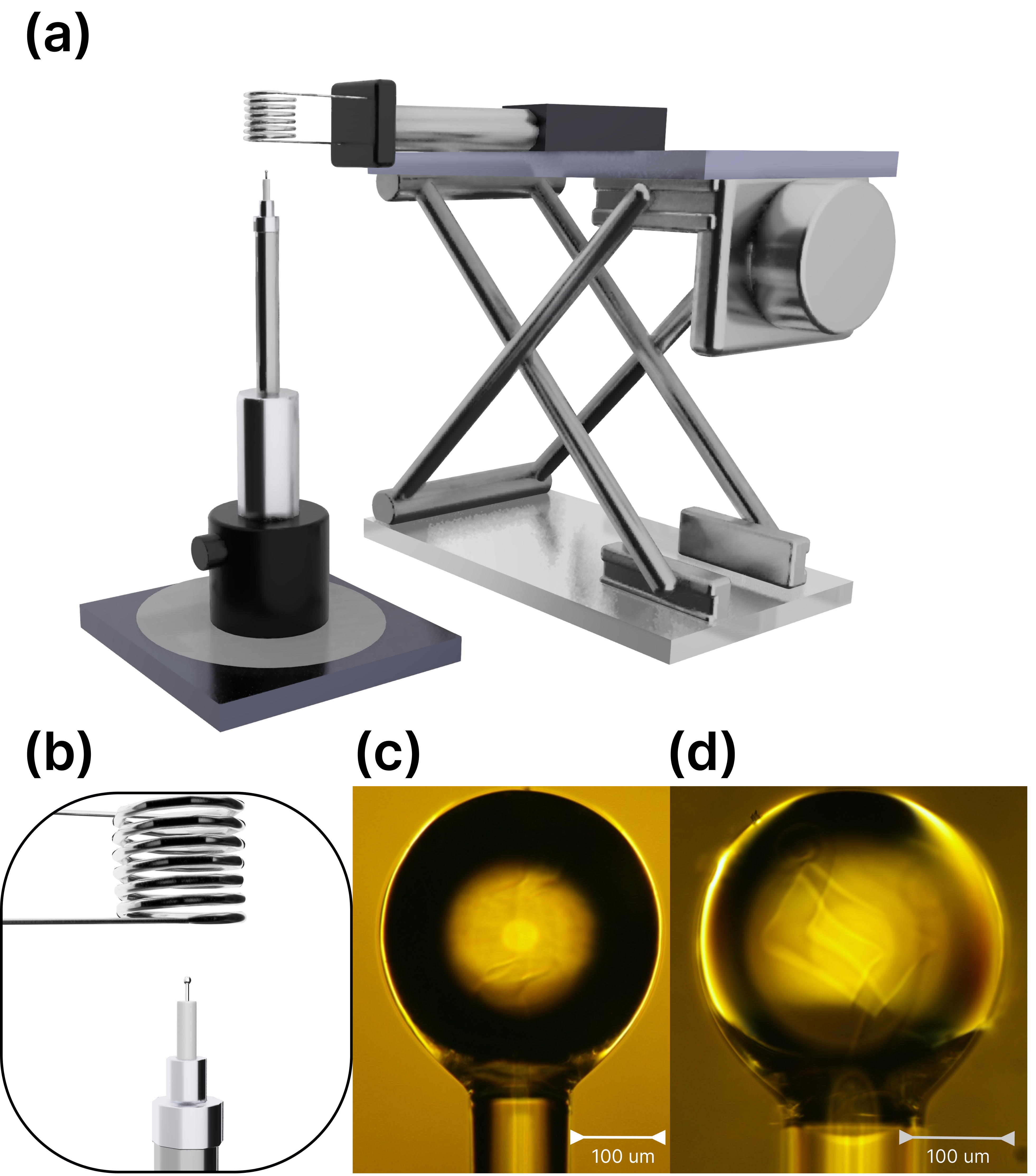}
\caption{(a) WGM ZBLAN microspherical resonator fabrication setup, (b) close-up view of the heating element (nichrome coil), (c) side view focused on the surface of a 300-$\mu$m-diameter ZBLAN microsphere, (d) side view, focused on center of the ZBLAN microsphere of a 270-$\mu$m-diameter. Blurred patterns in the center of the microsphere refers to the thermally modified core of the respective ZBLAN fiber}
\label{fig:fabrication}
\end{figure}

ZBLAN WGM microresonators were made by heating the tip of the corresponding optical fiber to the melting temperature, as a result the molten fiber tip acquired a spherical shape because of the surface tension \cite{way_high-q_2012}. 
Special aspect of the manufacturing of WGM ZBLAN microresonators is a slight difference in the glass transition and melting temperatures of the ZBLAN. The glass-forming ability depends on the resistance of the glass to the crystallization process when cooled in the temperature range between the melting point and the crystallization temperature. For ZBLAN the difference between these temperatures is relatively small - the crystallization temperature varies in the range of 350-400 \textdegree C, and the melting temperature is $\sim$ 450 \textdegree C. This can lead to the fact that during heating and cooling, the number of scattering centers on the surface and inside the resonator increases, which negatively affects the quality factor due to scattering. For multicomponent glasses such as ZBLAN the number of scattering centers can be even bigger due to the formation of different types of crystalline phases \cite{ong_suppression_2019}. 
It was shown in \cite{ong_suppression_2019} that crystallization in ZBLAN glass is highly dependent on the rate of heating and cooling of the glass. This effect can be used to decrease the crystallization. Non-uniform temperature distribution during heating on the glass can also adversely affects the quality factor \cite{way_high-q_2012}. The lack of symmetry of the heating element can lead to both an increase in the number of the crystallization centers and deformation of the spherical shape of the resonator. 
High-Q-factor WGM microresonators set requirements for the quality of manufacturing; in particular, the quality factor can be affected by shape and surface defects, as well as inhomogeneities on the subsurface layer of the microresonator \cite{grudinin_ultra_2006, gorodetsky_ultimate_1996} or suffered from crystallization \cite{ong_suppression_2019}.
In order to avoid the above problems, we have optimized the manufacturing process. For fabrication ZBLAN WGM microresonators we used modified method described in the work \cite{way_high-q_2012}. \textcolor{black}{The modification of the fabrication process mainly refers to the values of temperature and duration of the heating }\textcolor{black}{and cooling. An overheating result in irreversible chemical reactions in the ZBLAN, while underheating or rapid temperature change produce inhomogeneities on the surface which leads to dramatic Q-factor degradation. Both processes are not amenable to simulation, so the optimization was made experimentally by step-by-step trial and error method.} \textcolor{black}{ It was revealed that proper choice of these parameters drastically enhances the quality of the manufactured microresonators and allows to obtain the absorption-limited Q-factor.}
Fig. \ref{fig:fabrication} represents the fabrication setup. As a material we used commercial step-index single-mode fluorozirconate (ZBLAN) fiber with 9/125 core/cladding diameters, NA = 0.19, attenuation < 0.2 dB/m (from 2.3 to 3.6 $\mu$m)) made by Thorlabs (Fiber Identification - IRZS23). The manufacturing process of microspherical resonators from optical fibers includes three main stages: preliminary fiber preparation, melting and cooling processes. Preliminary fiber preparation includes cleaning the piece of fiber (2-3 cm) from the protective polymer coating with acetone, cutting off the tips of the fiber with fiber splicer, cleaning it with isopropanol. This preliminary preparation was of special importance for achieving high Q-factor. Then cleaned piece of fiber fixed into a ferrule of the FC/APC SM fiber connector with epoxy UV glue. After first stage of preparation the fiber connector with the piece of prepared fiber is fixed on a optical post under the heating element along its symmetry axis. A nichrome (wire diameter 0.16 mm) coil with a diameter of 2 mm and a height of 3 mm is used as a heating element to obtain a symmetrical temperature distribution. Using precise XYZ translational stage the heating element is placed to the required position - the tip of the fiber is supposed to be placed in the center of the heater. The experimental selection of the appropriate parameters gives us the next values of temperature and duration of the heating. The heating element is powered with a power supply with the parameters V = 6.9V, I = 1.4A with the overall heating time of 1-2 seconds, which gives central temperature inside the heating element 500 \textdegree C. Calibration of the temperature is carried out before the manufacturing process using a thermocouple, placed in the center of the heating element. The final diameter of the microsphere depends on the length of the fiber that placed inside the heating element before heating. We place the tip of the fiber at length from 0.2 to 0.4 mm from the bottom edge of the coil to vary the resulting diameters of the microspheres from 2-2.5 diameter of the fiber cladding. During melting and sphere formation process, the length of the fiber decreases and comes out of the heating element where the microsphere naturally cooled down \textcolor{black}{at room temperature (22 \textdegree C)}. As a result, ZBLAN microspheres with diameters of $\sim$250–350 $\mu$m are fabricated.

After cooling process the microsphere's shape and surface quality control is carried out with microscope as the first step of quality control. The photographs of two samples with the diameters of 270 and 300 $\mu$m are shown in the Fig. \ref{fig:fabrication} (c,d). One may see blurry patterns inside the microsphere in the Fig. \ref{fig:fabrication} (c,d), which are related to the deformed during the fabrication process core. The core has slightly different refractive index and ZBLAN is transparent in visible band, so that we are able to see internal structure of the microsphere. Since WGMs are localized near the surface, the core does not affect the Q-factor. As a result of the analysis of the obtained experimental data, we can point that the key stage in the manufacturing process is the preliminary fiber preparation. \textcolor{black}{With a clean surface of the fiber, the fabrication method with the proper parameters shows a high reproducibility with constant controlled parameters of diameters and surface quality.} The second step of quality control is the WGM excitation with fabricated microspheres in the near-IR, described below. 

\section{Experimental setup}
\begin{figure}[htbp!]
\centering
\includegraphics[width=0.73\linewidth]{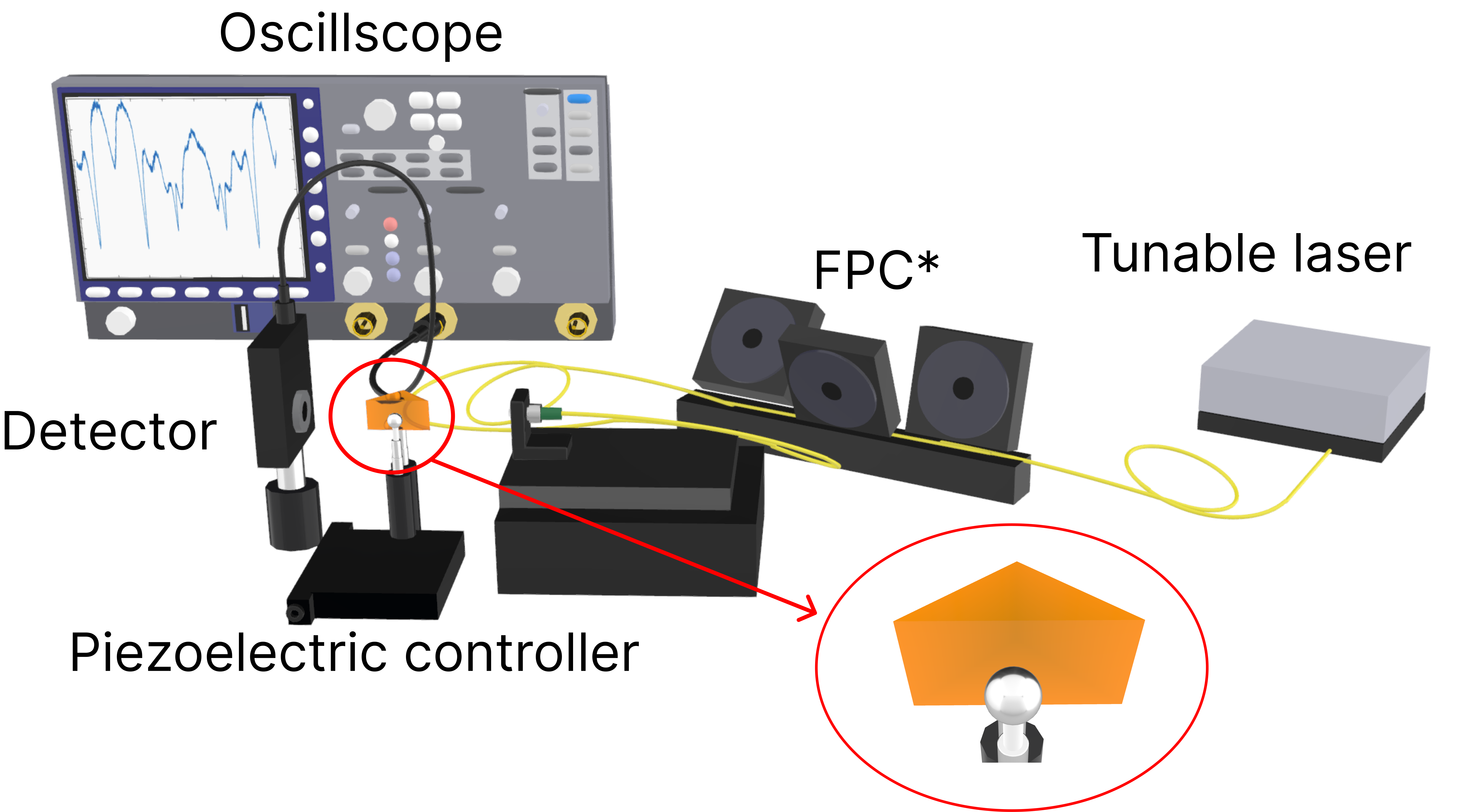}
\caption{Schematic illustration of the experimental setup. (FPC* (fiber polarization controller) is used only for excitation of the WGM in the near-IR range)}
\label{fig:setup}
\end{figure}

Fig. \ref{fig:setup} represents the general scheme of the experimental setup for near- and mid-IR characterization of ZBLAN microspheres.
 
A tunable 1.55 $\mu$m continuous wave laser diode pig-tailed with optical fiber was used to excite WGM \textcolor{black}{(in-coupled power is up to 0.5 mW)}. After polarization controller (FPC) light is emitted to the free space and coupled to the microresonator through the collimating lens (CaF$_2$) and the coupling element. 
The microresonator is mounted on XYZ translational stage with piezo controller. The transmitted light is collected on the photodetector (Ge biased detector by Thorlabs). A microscope was used to monitor the alignment of the microresonator. The frequency was calibrated using a Fabry-Pérot silica etalon (0.05 $cm^{-1}$). The choice of appropriate transparent materials in order to satisfy the conditions of phase matching for 1.55 and 2.64 $\mu$m wavelengths with ZBLAN microspheres is limited. After calculations of the excitation angle for different prism materials, we chose the ZnSe triangular prism as a coupling element.

 \section{Near-IR Q-factor measurement}
 
\begin{figure}[htbp!]
\centering
\includegraphics[width=0.68\linewidth]{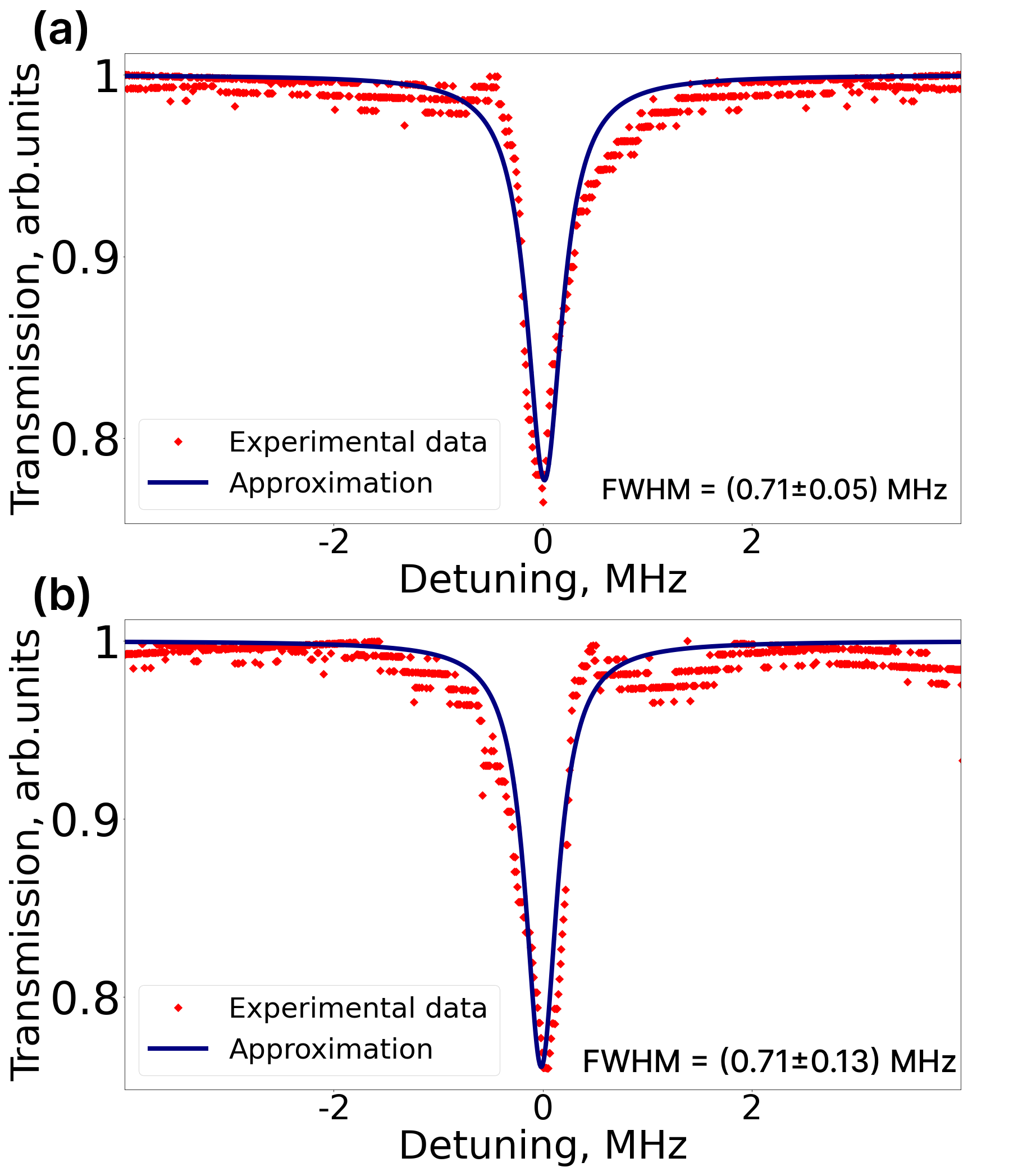}
\caption{Transmission spectrum of the microsphere in critically coupling regime upon forward (a) and backward (b) scan (Q-factor is ($2.7\pm0.2)\cdot10^8$). Blue curve corresponds to the Lorentzian fit.}
\label{fig:1550}
\end{figure}
\textcolor{black}{For microresonator characterization we measured transmission spectra sweeping probe laser frequency and many different families were observed.
 } To calculate the quality factor of the excited WGM the full width at half maximum (FWHM) of the resonance curve was measured. Quality factor measured in the critical coupling regime provide information about the intrinsic Q-factor. Varying the distance between the coupler and microresonator one may change the microresonator loading (coupling efficiency). \textcolor{black}{Maximum transmission dip appears at critical coupling where losses due to the coupling are equal to the intrinsic losses of the microresonator \cite{Gorodetsky1994high}.} 
For a microsphere with a diameter of 310 $\mu$m, we measured transmission spectra for different gaps from 0 to 1.5 microns between coupler prism and microsphere corresponding to different loading, from undercoupled to overcoupled regimes. At the critical coupling the loaded quality factor Q was $(2.7\pm0.2)\cdot10^8$, which leads to intrinsic Q = $(5.4\pm0.4)\cdot10^8$.
According to manufacturer specifications optical losses (for single mode ZBLAN optical fiber IRZS23 from Thorlabs) at 1.55 $\mu$m are equal to $\alpha = 0.1\, dB/m$ which corresponds to $Q = 5\cdot10^8$. Therefore, we reached the limit of Q defined by material absorption. The overcoupled Q at zero gap for the same mode was additionally measured as $(8.6\pm0.3)\cdot10^6$. \textcolor{black}{For all fabricated microresonators with different diameters (250-350 $\mu$m) the values of the intrinsic Q-factor were the same.} Transmission spectrum of the microsphere in critically coupling regime upon forward and backward frequency scanning at 1.55 $\mu$m is shown in Fig. \ref{fig:1550}(a,b). WGM at forward and backward frequency scans have the same linewidth so it indicates that there is no nonlinearity and it is appropriate to approximate the resonance curve with the Lorentz fit. Blue curve in Fig. \ref{fig:1550} corresponds to Lorentzian approximation. \textcolor{black}{The slight difference between the transmission and approximation lines are caused by laser power fluctuations and parasitic Fabry-Pérot interference.} Coupling efficiency was about 20$\%$.

\section{Mid-IR Q-factor measurement}

\begin{figure}[htbp!]
\centering
\includegraphics[width=0.99\linewidth]{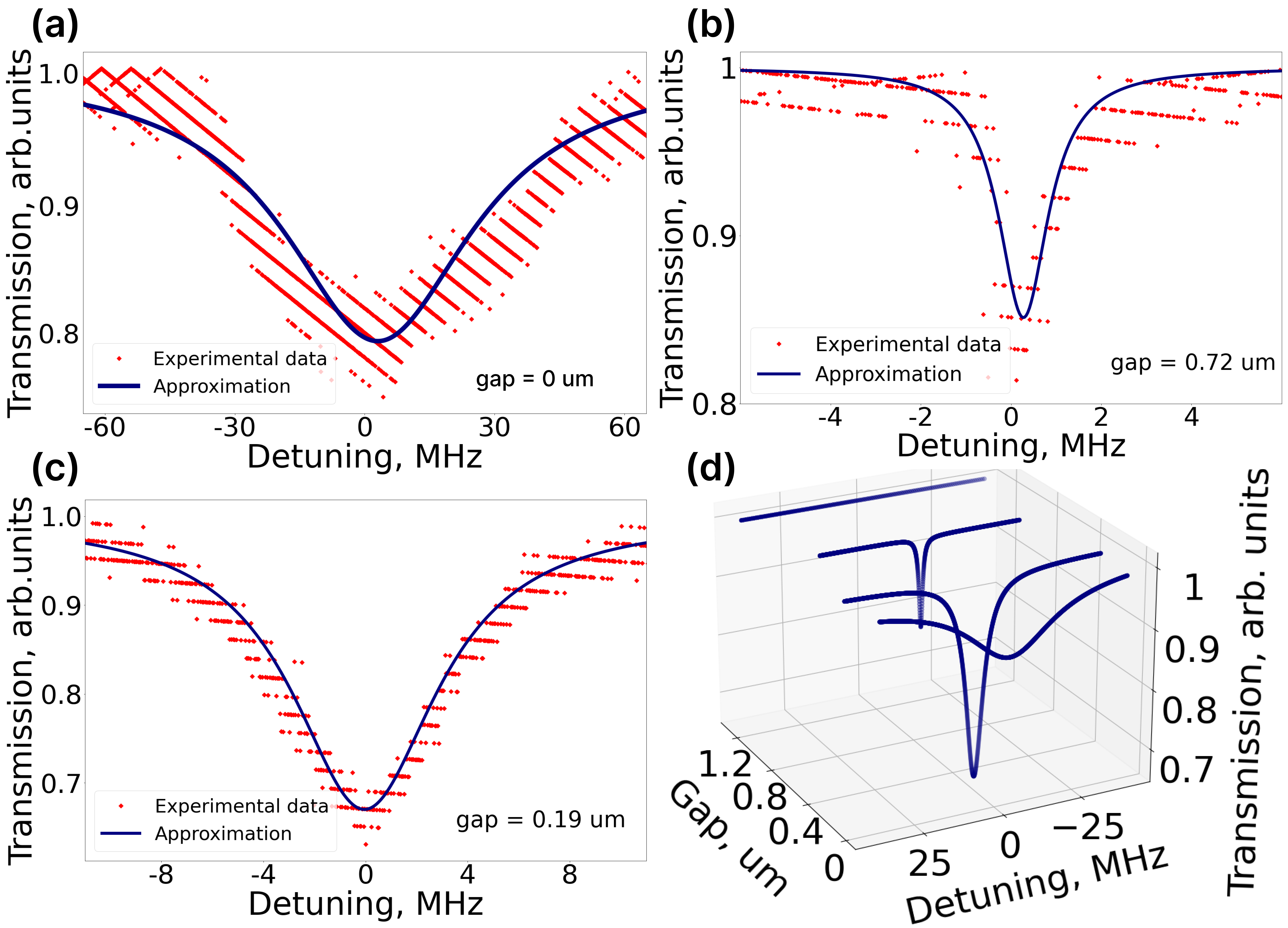}
\caption{\textcolor{black}{Transmission spectrum of the microsphere depending on distance between ZnSe prism and ZBLAN microsphere at 2.64 $\mu$m: (a) gap= 0.72 $\mu$m, Q= $(1.13\pm0.22)\cdot10^8$, (b) gap= 0.19 $\mu$m, Q= $(1.8\pm0.1)\cdot10^7$, (c) gap= 0 $\mu$m, Q= $(2.7\pm0.2)\cdot10^6$. Blue curve corresponds to Lorentzian fitting. (d) Transmission spectrum as a function of gap between prism and microsphere.} 
}
\label{fig:2640}
\end{figure}

For mid-IR characterization of the ZBLAN microspheres we used the same experimental setup (Fig. \ref{fig:setup}) without FPC. A tunable 2.64 $\mu$m continuous wave DFB diode laser from Nanoplus \textcolor{black}{(in-coupled laser power is up to 0.25 mW)} emitted to the free space and through the collimating lens system fed into a single-mode ZrF$_4$ fiber. After ZrF$_4$ fiber light is emitted to the free space and coupled to the microresonator through the collimating lens (CaF$_2$) and the triangular prism (ZnSe). The radiation transmitted through the coupling element is collected by a photodetector (InAs by IoffeLED). The frequency scanning was calibrated using a Fabry-Pérot germanium etalon (0.05 $cm^{-1}$).
For mid-IR characterization we measured Q-factors for the same microsphere in both undercoupled and overcoupled regime depending on the distance between ZnSe prism and ZBLAN microsphere. As a result, the quality factor Q in the overcoupled regime was $(2.7\pm0.2)\cdot10^6$), in the undercoupled - at gap of 0.19 $\mu$m Q = $(1.8\pm0.1)\cdot10^7$ and at gap - 0.72 $\mu$m Q = $(1.13\pm0.22)\cdot10^8$. Fig. \ref{fig:2640} represents the transmission spectrum depending on gap. From 20 $\%$ to 30 $\%$ of light was coupled to the microsphere depending on gap. \textcolor{black}{The coefficient of determination for the Lorentzian approximation} (blue curve) was also over 0.9 for Fig. \ref{fig:2640}(a,c). For Fig. \ref{fig:2640}(b) approximation line shows a good fitting, but \textcolor{black}{the R-squared factor} is only 0.8. This might be explained by the resolution limit of the \textcolor{black}{laser source linewidth (< 3 MHz)}. 
The minimum of the material losses in ZBLAN optical fiber is around 2-3 $\mu$m according to the calculated absorption-limited Q-factor as function of wavelength based on values provided by Thorlabs for ZBLAN fiber. However, the quality factor measured at 2.64 $\mu$m is a bit lower than at 1.55 $\mu$m apparently due to the resolution limit of the laser source linewidth.

\section{Conclusion}

\textcolor{black}{In conclusion, we have demonstrated a fabrication technique of the high-Q ZBLAN microspheres with the diameters of 250 to 350 $\mu$m based on melting of commercially available optical fiber. It is revealed that careful preliminary blank preparation and accurate choice of the heating process parameters provides significant enhancement of the microresonator quality factor.} \textcolor{black}{This is quite the opposite to the manufacturing of the widely used fused silica resonators, for which it is enough to exceed the melting temperature while the exact temperature value, and the rate of heating and cooling do not matter.} Fabricated ZBLAN microspheres show quality factor values close to the absorption-limited Q-factor at 1.55 $\mu$m (intrinsic Q-factor = $(5.4\pm0.4)\cdot10^8$, $\alpha = 2.5\, cm^{-1}$). We effectively excited WGM in the mid-IR range and demonstrated quality factor of $(1.13\pm0.22)\cdot10^8$ at 2.64 $\mu$m. \textcolor{black}{The threshold for nonlinear effects such as frequency comb generation decreases proportionally to the square of the quality factor, so increasing it by a factor of 15 provides more than 200 times reduction of the required pump power.}
The fabrication method can be applied for producing the rare-earth-doped ZBLAN WGM microresonators. The demonstration of ultra-high Q-factors in ZBLAN paves the way for constructing microlasers based on such microspheres and of special interest for biosensing applications in the mid-IR \textcolor{black}{and may encourage careful investigation of other glassy materials for microresonators.} 

\begin{backmatter}
\bmsection{Funding}
The work was supported by the Russian Foundation for Basic Research (project No 20-32-90184). 


\bmsection{Disclosures}
The authors declare no conflicts of interest.
\end{backmatter}

\bibliography{ZBLAN_article}

\begin{thebibliography}{10}
\newcommand{\enquote}[1]{``#1''}

\bibitem{braginsky_quality-factor_1989}
V.~B. Braginsky, M.~L. Gorodetsky, and V.~S. Ilchenko, \enquote{Quality-factor
  and nonlinear properties of optical whispering-gallery modes,}
  {\protect\JournalTitle{Physics Letters A}} \textbf{137}, 393--397 (1989).

\bibitem{vahala_optical_2003}
K.~J. Vahala, \enquote{Optical microcavities,} {\protect\JournalTitle{Nature}}
  \textbf{424}, 839–846 (2003).

\bibitem{Ilchenko2006-II}
V.~Ilchenko and A.~Matsko, \enquote{Optical resonators with whispering-gallery
  modes-part ii: applications,} {\protect\JournalTitle{IEEE Journal of Selected
  Topics in Quantum Electronics}} \textbf{12}, 15--32 (2006).

\bibitem{Zhu2010}
J.~Zhu, S.~K. Ozdemir, Y.-F. Xiao, L.~Li, L.~He, D.-R. Chen, and L.~Yang,
  \enquote{On-chip single nanoparticle detection and sizing by mode splitting
  in an ultrahigh-{Q} microresonator,} {\protect\JournalTitle{Nature
  Photonics}} \textbf{4}, 46--49 (2010).

\bibitem{Ward2011}
J.~Ward and O.~Benson, \enquote{{WGM} microresonators: sensing, lasing and
  fundamental optics with microspheres,} {\protect\JournalTitle{Laser \&
  Photonics Reviews}} \textbf{5}, 553--570 (2011).

\bibitem{Lin:17}
G.~Lin, A.~Coillet, and Y.~K. Chembo, \enquote{Nonlinear photonics with
  high-{Q} whispering-gallery-mode resonators,} {\protect\JournalTitle{Adv.
  Opt. Photon.}} \textbf{9}, 828--890 (2017).

\bibitem{Kippenberg2018}
T.~J. Kippenberg, A.~L. Gaeta, M.~Lipson, and M.~L. Gorodetsky,
  \enquote{Dissipative {K}err solitons in optical microresonators,}
  {\protect\JournalTitle{Science}} \textbf{361}, eaan8083 (2018).

\bibitem{PASQUAZI20181}
A.~Pasquazi, M.~Peccianti, L.~Razzari, D.~J. Moss, S.~Coen, M.~Erkintalo, Y.~K.
  Chembo, T.~Hansson, S.~Wabnitz, P.~Del’Haye, X.~Xue, A.~M. Weiner, and
  R.~Morandotti, \enquote{Micro-combs: A novel generation of optical sources,}
  {\protect\JournalTitle{Physics Reports}} \textbf{729}, 1--81 (2018).

\bibitem{deng_demonstration_2014}
Y.~Deng, R.~K. Jain, and M.~Hossein-Zadeh, \enquote{Demonstration of a cw room
  temperature mid-{IR} microlaser,} {\protect\JournalTitle{Optics Letters}}
  \textbf{39}, 4458 (2014).

\bibitem{behzadi_power_2018}
B.~Behzadi, R.~K. Jain, and M.~Hossein-Zadeh, \enquote{Power scaling of
  narrow-linewidth mid-{IR} spherical microlasers,}
  {\protect\JournalTitle{Laser Physics Letters}} \textbf{15}, 085112 (2018).

\bibitem{he_whispering_2013}
L.~He, c.~K. \"{O}zdemir, and L.~Yang, \enquote{Whispering gallery microcavity
  lasers: {WGM} microlasers,} {\protect\JournalTitle{Laser \& Photonics
  Reviews}} \textbf{7}, 60--82 (2013).

\bibitem{Liang2015}
W.~Liang, V.~S. Ilchenko, D.~Eliyahu, A.~A. Savchenkov, A.~B. Matsko,
  D.~Seidel, and L.~Maleki, \enquote{Ultralow noise miniature external cavity
  semiconductor laser,} {\protect\JournalTitle{Nature Communications}}
  \textbf{6}, 7371 (2015).

\bibitem{Galiev:18}
R.~R. Galiev, N.~G. Pavlov, N.~M. Kondratiev, S.~Koptyaev, V.~E. Lobanov, A.~S.
  Voloshin, A.~S. Gorodnitskiy, and M.~L. Gorodetsky, \enquote{Spectrum
  collapse, narrow linewidth, and {B}ogatov effect in diode lasers locked to
  high-{Q} optical microresonators,} {\protect\JournalTitle{Opt. Express}}
  \textbf{26}, 30509--30522 (2018).

\bibitem{shitikov2021self}
A.~E. Shitikov, V.~E. Lobanov, N.~M. Kondratiev, A.~S. Voloshin, E.~A.
  Lonshakov, and I.~A. Bilenko, \enquote{Self-injection locking of a
  gain-switched laser diode,} {\protect\JournalTitle{Physical Review Applied}}
  \textbf{15}, 064066 (2021).

\bibitem{yu_mode-locked_2016}
M.~Yu, Y.~Okawachi, A.~G. Griffith, M.~Lipson, and A.~L. Gaeta,
  \enquote{Mode-locked mid-infrared frequency combs in a silicon
  microresonator,} {\protect\JournalTitle{Optica}} \textbf{3}, 854 (2016).

\bibitem{griffith_silicon_2015}
A.~G. Griffith, R.~K. Lau, J.~Cardenas, Y.~Okawachi, A.~Mohanty, R.~Fain,
  Y.~H.~D. Lee, M.~Yu, C.~T. Phare, C.~B. Poitras, A.~L. Gaeta, and M.~Lipson,
  \enquote{Silicon-chip mid-infrared frequency comb generation,}
  {\protect\JournalTitle{Nature Communications}} \textbf{6}, 6299 (2015).

\bibitem{shitikov2020microresonator}
A.~E. Shitikov, O.~V. Benderov, N.~M. Kondratiev, V.~E. Lobanov, A.~V. Rodin,
  and I.~A. Bilenko, \enquote{Microresonator and laser parameter definition via
  self-injection locking,} {\protect\JournalTitle{Physical Review Applied}}
  \textbf{14}, 064047 (2020).

\bibitem{shankar_integrated_2013}
R.~Shankar, I.~Bulu, and M.~Lončar, \enquote{Integrated high-quality factor
  silicon-on-sapphire ring resonators for the mid-infrared,}
  {\protect\JournalTitle{Applied Physics Letters}} \textbf{102}, 051108 (2013).

\bibitem{klitzing_frequency_2001}
W.~Klitzing, R.~Long, V.~S. Ilchenko, J.~Hare, and V.~Lefèvre-Seguin,
  \enquote{Frequency tuning of the whispering-gallery modes of silica
  microspheres for cavity quantum electrodynamics and spectroscopy,}
  {\protect\JournalTitle{Optics Letters}} \textbf{26}, 166--168 (2001).

\bibitem{armani_ultra-high-q_2003}
D.~K. Armani, T.~J. Kippenberg, S.~M. Spillane, and K.~J. Vahala,
  \enquote{Ultra-high-{Q} toroid microcavity on a chip,}
  {\protect\JournalTitle{Nature}} \textbf{421}, 925--928 (2003).

\bibitem{anashkina_optical_2021}
E.~A. Anashkina, M.~P. Marisova, T.~Salgals, J.~Alnis, I.~Lyashuk, G.~Leuchs,
  S.~Spolitis, V.~Bobrovs, and A.~V. Andrianov, \enquote{Optical {Frequency}
  {Combs} {Generated} in {Silica} {Microspheres} in the {Telecommunication}
  {C}-, {U}-, and {E}-{Bands},} {\protect\JournalTitle{Photonics}} \textbf{8},
  345 (2021).

\bibitem{anashkina_kerr_2021}
E.~A. Anashkina, V.~Bobrovs, T.~Salgals, I.~Brice, J.~Alnis, and A.~V.
  Andrianov, \enquote{Kerr {Optical} {Frequency} {Combs} {With} {Multi}-{FSR}
  {Mode} {Spacing} in {Silica} {Microspheres},} {\protect\JournalTitle{IEEE
  Photonics Technology Letters}} \textbf{33}, 453--456 (2021).

\bibitem{shen_observation_2015}
Z.~Shen, Z.-H. Zhou, C.-L. Zou, F.-W. Sun, G.-P. Guo, C.-H. Dong, and G.-C.
  Guo, \enquote{Observation of high-{Q} optomechanical modes in the mounted
  silica microspheres,} {\protect\JournalTitle{Photonics Research}} \textbf{3},
  243--247 (2015).

\bibitem{ilchenko_nonlinear_2004}
V.~S. Ilchenko, A.~A. Savchenkov, A.~B. Matsko, and L.~Maleki,
  \enquote{Nonlinear optics and crystalline whispering gallery mode cavities,}
  {\protect\JournalTitle{Phys. Rev. Lett.}} \textbf{92}, 043903 (2004).

\bibitem{Fujii:20}
S.~Fujii, Y.~Hayama, K.~Imamura, H.~Kumazaki, Y.~Kakinuma, and T.~Tanabe,
  \enquote{All-precision-machining fabrication of ultrahigh-{Q} crystalline
  optical microresonators,} {\protect\JournalTitle{Optica}} \textbf{7},
  694--701 (2020).

\bibitem{gorodetsky_ultimate_1996}
M.~L. Gorodetsky, A.~A. Savchenkov, and V.~S. Ilchenko, \enquote{Ultimate {Q}
  of optical microsphere resonators,} {\protect\JournalTitle{Optics Letters}}
  \textbf{21}, 453--455 (1996).

\bibitem{miya1979ultimate}
T.~Miya, Y.~Terunuma, T.~Hosaka, and T.~Miyashita, \enquote{Ultimate low-loss
  single-mode fibre at 1.55 $\mu$m,} {\protect\JournalTitle{Electronics
  Letters}} \textbf{4}, 106--108 (1979).

\bibitem{poulain_verres_1975}
M.~Poulain, M.~Poulain, and J.~Lucas, \enquote{Verres fluores au tetrafluorure
  de zirconium proprietes optiques d'un verre dope au {N}d$^{3+}$,}
  {\protect\JournalTitle{Materials Research Bulletin}} \textbf{10}, 243--246
  (1975).

\bibitem{zhang1994evaluation}
L.~Zhang, F.~Gan, and P.~Wang, \enquote{Evaluation of refractive-index and
  material dispersion in fluoride glasses,} {\protect\JournalTitle{Applied
  optics}} \textbf{33}, 50--56 (1994).

\bibitem{harrington_2004}
J.~A. Harrington, \emph{Infrared fibers and their applications}, SPIE Press
  monograph; PM135 (SPIE Press, 2004).

\bibitem{parker_fluoride_1989}
J.~M. Parker, \enquote{Fluoride {Glasses},} {\protect\JournalTitle{Annual
  Review of Materials Science}} \textbf{19}, 21--41 (1989).

\bibitem{way_high-q_2012}
B.~Way, R.~K. Jain, and M.~Hossein-Zadeh, \enquote{High-{Q} microresonators for
  mid-{IR} light sources and molecular sensors,} {\protect\JournalTitle{Optics
  Letters}} \textbf{37}, 4389--4391 (2012).

\bibitem{saad_fluoride_2009}
M.~Saad, \enquote{Fluoride glass fiber: State of the art,}
  {\protect\JournalTitle{Proc {SPIE}}} \textbf{7316} (2009).

\bibitem{wetenkamp_optical_1992}
L.~Wetenkamp, G.~West, and H.~Többen, \enquote{Optical properties of rare
  earth-doped {ZBLAN} glasses,} {\protect\JournalTitle{Journal of
  Non-Crystalline Solids}} \textbf{140}, 35--40 (1992).

\bibitem{ong_suppression_2019}
T.-C. Ong, B.~Fogarty, T.~Steinberg, E.~Jaatinen, and J.~Bell,
  \enquote{Suppression of crystallization in {ZBLAN} glass by rapid heating and
  cooling processing,} {\protect\JournalTitle{International Journal of Applied
  Glass Science}} \textbf{10}, 391--400 (2019).

\bibitem{grudinin_ultra_2006}
I.~S. Grudinin, A.~B. Matsko, A.~A. Savchenkov, D.~Strekalov, V.~S. Ilchenko,
  and L.~Maleki, \enquote{Ultra high {Q} crystalline microcavities,}
  {\protect\JournalTitle{Optics Communications}} \textbf{265}, 33--38 (2006).

\bibitem{Gorodetsky1994high}
M.~Gorodetsky and V.~Ilchenko, \enquote{High-{Q} optical whispering-gallery
  microresonators: precession approach for spherical mode analysis and emission
  patterns with prism couplers,} {\protect\JournalTitle{Optics Communications}}
  \textbf{113}, 133--143 (1994).

\end{thebibliography}

\bibliographyfullrefs{ZBLAN_article}


\end{document}